\documentclass[aps,twocolumn]{revtex4}
\usepackage[dvips]{graphicx}
\def\be{\begin{equation}}
\def\ee{\end{equation}}
\def\bea{\begin{eqnarray}}          
\def\eea{\end{eqnarray}}
\def\bi{\begin{itemize}}
\def\ei{\end{itemize}}

\begin{document}

\title{Self-localized impurities embedded in a one dimensional 
 Bose-Einstein condensate and their quantum fluctuations}

\author{Krzysztof Sacha$^{1,2}$ and Eddy Timmermans$^1$}

\affiliation{$^1$ T-4, Theory Division, Los Alamos National Laboratory, Los Alamos, NM
87545 \\
$^2$ Institute of Physics, Jagiellonian University,
          Reymonta 4, 30-059 Krak\'ow, Poland  }

\begin{abstract}
We consider the self-localization of neutral impurity atoms in a 
Bose-Einstein condensate in a 1D model. Within the strong coupling
approach, we show that the self-localized state exhibits parametric soliton
behavior.  The corresponding stationary states are analogous to the solitons of non-linear
optics and to the solitonic solutions of the Schr\"odinger-Newton
equation (which appears in models that consider the connection between quantum
mechanics and gravitation).  In addition, we present a Bogoliubov-de~Gennes
formalism to describe the quantum fluctuations around the product state 
of the strong coupling description.  Our fluctuation calculations yield
the excitation spectrum and reveal considerable corrections to the strong 
coupling description.  The knowledge of the spectrum allows a 
spectroscopic detection of the impurity self-localization phenomenon.
\end{abstract}

\maketitle

PACS 03.75.Hh, 67.40.Yv

%%%%%%%%%%%%%%%%%%%%%%%%%%%%%%%%%%%%%%%%%%%%%%%%%%%%%%%%%%%%%%%%%%%%%
\section{Introduction}

Recent work pointed out that neutral impurity atoms immersed in a dilute
gas Bose-Einstein condensate (BEC) can spontaneously self-localize
\cite{eddy,blume}. Repeated measurements of the impurity position 
would observe this particle's wavefunction to have a finite extent, even if its
average position cannot be predicted a priori (at least if the
BEC were homogeneous in the absence of the impurity).  In three
dimensions (3D), the BEC-impurity physics is akin to that of an electron moving in
a polar crystal (a polaron) \cite{Mahan}. In 3D, the self-localization of a neutral
atom BEC-impurity occurs when the 
magnitude of the impurity-boson scattering length exceeds a minimal
value (regardless of the sign of that interaction).  The self-localized
BEC impurity state resembles that of a small polaron.  Traditional
(i.e. electronic) small polarons have been described successfully in the
strong coupling limit by the Landau-Pekar  treatment \cite{landau}
which assumes a wavefunction that is the product of a single electron wavefunction
and a bosonic phonon state.  The self-localized BEC-impurity
has been described similarly [1,2] by the product of a single particle impurity
wavefunction and a BEC-state.  Linearizing the BEC-response 
to the impurity density then gives an effective impurity equation 
of a single particle experiencing an attractive Yukawa self-interaction potential [1].  While
the ensuing analysis is elegant, the validity of the starting point 
(i.e. the product state) is much  more questionable than in the description 
of traditional polarons.   In the latter case, the large mass difference of 
crystal ions and electrons implies a clear separation of time scales
which justifies the product state -- one can always think
of the slow (boson) field adjusting itself to the time averaged field
of the fast (electron) particle.  In the cold atom BEC-impurity,
the boson and impurity atoms tend to have similar masses
and the time scales do not separate.  Hence, we can expect the deviations
from the product state, the fluctuations that describe the 
entanglement of the impurity and boson degrees of freedom, to
become much more significant.  

In this paper, we describe
the one-dimensional analogue of the BEC-impurity, realizable
in quasi-one-dimensional BEC's.  For this system and for a
specific choice of the parameters, we find an exact
solution to the strong coupling equations and we 
solve for the quantum
fluctuations as a function of the impurity-boson mass ratios.  The
explicit solution to the strong-coupling equations are parametric solitons,
which establishes that the 1D analogues of self-localized BEC-impurities 
are optical solitons.   Specifically, the impurity soliton solutions  are 
resemble the solitons that appear when
a quadratic nonlinearity is responsible for second-harmonic 
generation \cite{drummond96,karpierz}. 
The Schr\"odinger-Newton equation
that models the gravitational interaction while preventing the
quantum-mechanical spreading  of the
center-of-mass position of macroscopic objects \cite{ruffini,diosi,penrose}, as well
as the mean-field equations of coupled atomic-molecular BEC's
\cite{drummondPRL98,drummondPRA04} possess similar parametric soliton 
solutions.  The connection with these
very different fields of physics further extends the scope of the
BEC-impurity physics.

The experimental realization of the BEC-impurity systems requires the creation of
distinguishable atoms in BEC's.  This feat has been realized by converting a fraction
of the  BEC-atoms with a two-photon Raman transition to a different spin 
state in order to observe the superfluid suppression of slow impurity 
scattering by the BEC \cite{ketterle1}, or by trapping distinct species
of atoms \cite{ketterle2,roati}
or isotopes \cite{truscott}.  
In this paper, we describe a BEC trapped in
a one dimensional box.  Our predictions apply to atomic traps with
strong confinement in the two transverse directions (quasi-1D).  A
1D box potential was recently achieved experimentally \cite{raizen}, but
our calculations should also describe the physics of quasi 1D-BEC's
with trapping potentials that vary slowly in the longitudinal direction.

The paper is organized as follows: In section II, we introduce the model
and work within the product state description.  Section III presents a
formalism that goes beyond the product state ansatz by means of
a Bogoliubov description of the quantum fluctuations.  In section IV,
we discuss our numerical results, and we conclude in section V.

%%%%%%%%%%%%%%%%%%%%%%%%%%%%%%%%%%%%%%%%%%%%%%%%%%%%%%%%%%%%%%%%%%%%%
\section{Strong coupling approach}
\label{strc}

We consider $M$ impurity bosonic atoms immersed in a homogeneous 
BEC in a 1D model. The Hamiltonian of the system reads
\bea
\hat H&=&\int {\rm d}x \left\{ \hat\varphi^\dagger(x)\left[
-\frac{\hbar^2\partial^2_x}{2m_{\rm B}}+\frac{\lambda_{{\rm BB}}}{2}
\hat\varphi^\dagger(x)\hat\varphi(x)
-\mu_{\rm B}\right]\hat\varphi(x) \right. \cr
&+&
\hat\psi^\dagger(x)\left[-\frac{\hbar^2\partial^2_x}{2m_{\rm I}}-E_{\rm I}\right]\hat\psi(x)
\cr 
&+&
\left.\lambda_{{\rm IB}}\hat\varphi^\dagger(x)\hat\varphi(x)\hat\psi^\dagger(x)\hat\psi(x)\right\},
\label{ham}
\eea
where $\hat{\phi}(x)$ and
$\hat{\psi}(x)$ stand for the condensate boson and impurity
atom field operators, respectively; $m_{\rm B}$ and $m_{\rm I}$ are the
boson and impurity masses, while $\mu_{\rm B}$ and $E_{\rm I}$
represent the chemical potentials of the BEC and impurity
systems. We assume contact boson-boson 
and impurity-boson interactions characterized by $\lambda_{{\rm BB}}$ and 
$\lambda_{{\rm IB}}$, respectively. We neglect the mutual interactions of impurity
atoms in the assumption that their number and local density remains 
sufficiently small.

In the strong coupling treatment \cite{landau,eddy}, we describe the system in terms
of a product state,
\be
\Psi
%(y_1,\dots,y_M,x_1,\dots,x_N)
=\eta(y_1)\dots\eta(y_M)\zeta(x_1)\dots\zeta(x_N),
\label{product}
\ee
where $\eta(x)$ is the wave-function occupied by the $M$ impurity atoms 
and $\zeta(x)$ is the single particle state occupied by the $N$ bosons. 
The expectation value 
of the Hamiltonian (\ref{ham}) for the product state (\ref{product})  
gives an energy that reaches its minimal value when the following 
equations are satisfied:
\bea
\left[-\frac{\hbar^2\partial^2_x}{2m_{\rm B}} +
\lambda_{{\rm BB}}\varphi^2(x)+\lambda_{{\rm IB}}\psi_0^2(x)\right]\varphi(x)=\mu_{\rm B}\varphi(x),
\cr
\left[-\frac{\hbar^2\partial^2_x}{2m_{\rm I}} +
\lambda_{{\rm IB}}\varphi^2(x)\right]\psi_0(x) =E_{\rm I}\psi_0(x),
\label{meanf1}
\eea
where 
\bea
\varphi(x)&=&\sqrt{N} \zeta(x), \cr
\psi_0(x)&=&\sqrt{M}\eta(x),
\eea
and we have assumed that both
$\varphi(x)$ and $\psi_0(x)$ are real-valued -- an assumption that is 
permitted since we wish to describe a ground state.  We also assume that 
$M$, the number of impurity atoms is small, so that we may expect only a 
slight modification of the boson wave-function $\varphi(x)$ with respect 
to the homogeneous BEC solution $\sqrt{\rho}$, where $\rho$ is the BEC 
density. Therefore we substitute
\be
\varphi(x) = \sqrt{\rho}+\phi_0(x),
\label{lin1}
\ee
into (\ref{meanf1}) and keep linear terms in $\phi_0(x)$ only,
\bea
\left[-\frac{\hbar^2\partial^2_x}{2m_{\rm B}} + 2\lambda_{{\rm BB}}\rho\right]\phi_0(x)
+\lambda_{{\rm IB}}\sqrt{\rho}\psi_0^2(x)=0, 
\label{meanf2a} 
\\
\left[-\frac{\hbar^2\partial^2_x}{2m_{\rm I}}+2\lambda_{{\rm IB}}\sqrt{\rho}\phi_0(x)\right]
\psi_0(x)=\tilde E_{\rm I}\psi_0(x), 
\label{meanf2b}
\eea
where $\mu_{\rm B}\approx \lambda_{{\rm BB}}\rho$ and 
$\tilde E_{\rm I}=E_{\rm I}-\lambda_{{\rm IB}}\rho$ \cite{eddy}.
This approximation is valid provided,
\be
\frac{\phi_0(x)}{\sqrt{\rho}}\ll 1.
\ee
The Eq.~(\ref{meanf2a}) can be solved in terms of the Green function of the 1D
Helmholtz equation,
\be
\phi_0(x)=-\frac{m_{\rm B}\lambda_{\rm IB}\sqrt{\rho}}{\hbar^2\chi}
\int {\rm d}y \psi_0^2(y)e^{-\chi|x-y|},
\label{green}
\ee
where
\be
\chi=\frac{2\sqrt{m_{\rm B}\lambda_{\rm BB}\rho}}{\hbar},
\ee
represents the BEC-coherence length.
The substitution of (\ref{green}) into Eq.~(\ref{meanf2b}) gives an equation
that describes impurity atoms self-interacting through an attractive
exponential potential. Note the difference with the 3D situation in which the 
self-interaction 
takes place through a Yukawa (screened Coulomb) potential \cite{eddy}.

Equations similar to Eqs.~(\ref{meanf2a},\ref{meanf2b}) occur in 
non-linear optics \cite{drummond96,karpierz}, in mean-field
descriptions of coupled atomic-molecular BEC's 
\cite{drummondPRL98,drummondPRA04} and in the Schr\"odinger-Newton
model \cite{diosi}.  These equations are known to possess parametric soliton
solutions.  In the present case, an analytical solution exists for a
particular value of the condensate density.  For 
\be
\rho=\frac{M^2m_{\rm I}^2\lambda_{{\rm IB}}^4}{36\hbar^2m_{\rm B}\lambda_{{\rm BB}}^3},
\label{rhos}
\ee
the Eqs.~(\ref{meanf2a},\ref{meanf2b}) transform into
\bea
\left[-\frac{\partial^2_x}{2}+\frac{M}{3}\right]\phi_0(x)+
\left(\frac{M}{6}\right)^{1/4}\sqrt{\frac{m_{\rm B}}{m_{\rm I}}}
{\rm sign}(\lambda_{\rm IB})\psi_0^2(x)=0, \cr
\left[-\frac{\partial^2_x}{2}+2\left(\frac{M}{6}\right)^{1/4}
\sqrt{\frac{m_{\rm I}}{m_{\rm B}}}{\rm sign}(\lambda_{\rm IB})
\phi_0(x)\right]\psi_0(x)=E\psi_0(x). \cr
&&
\label{meanf3}
\eea
These coupled equations possess a solitonic solution,
\bea
E&=&-\frac{M}{3}, \cr
\phi_0(x)&=&-{\rm sign}(\lambda_{\rm IB})\sqrt{\frac{3M^{3/2}m_{\rm B}}{8\sqrt{6}m_{\rm I}}}
\cosh^{-2}\left(\sqrt{\frac{M}{6}}x\right), \cr
\psi_0(x)&=&\sqrt{\frac{3M^{3/2}}{4\sqrt{6}}}
\cosh^{-2}\left(\sqrt{\frac{M}{6}}x\right).
\label{soliton}
\eea
In Eqs.~(\ref{meanf3},\ref{soliton}) we have scaled energy and length 
by $E_0$ and $x_0$ where
\bea
E_0&=&\frac{\lambda_{{\rm IB}}^2}{\hbar}\sqrt{\frac{m_{\rm B}\rho}{\lambda_{{\rm BB}}}}, \cr
x_0&=&\left(
\frac{\hbar^6\lambda_{{\rm BB}}}{m_{\rm I}^2m_{\rm B}\rho\lambda_{{\rm IB}}^4}
\right)^{1/4}.
\label{units}
\eea
The soliton (\ref{soliton}) describes a bound state of the impurity-BEC system
with  an impurity extend of order $\sqrt{6/M}$, which is precisely 
the BEC healing length since 
\be
\frac{\xi}{x_0}=\frac{\hbar}{x_0\sqrt{m_{\rm B}\lambda_{\rm
BB}\rho}}=\sqrt{\frac{6}{M}}.
\ee 
The self-localization takes place for either sign of the impurity-BEC
interaction ($\lambda_{\rm IB} >0$ and $\lambda_{\rm IB}<0$).
In addition to and caused by the impurity localization, the
condensate is deformed.  The BEC exhibits a dip in the density if the
impurity
and BEC-particle mutually repel ($\lambda_{\rm IB}>0$) and a hump if the
they attract ($\lambda_{\rm IB}<0$).

We have shown there exists a class of analytical solutions for a
particular choice of parameters of the 1D BEC-impurity system.
In general, one can solve the coupled equations
numerically.  Using the Gaussian ansatz for $\psi_{0}(x)$, one
can also show that there is impurity self-localization in the 1D-model
even as $\lambda_{\rm IB} \rightarrow 0$ \cite{fernando}.  This is
markedly different from the 3D-situation for which the impurity-BEC 
interaction must be sufficiently strong before self-localization sets 
in \cite{eddy}.

%%%%%%%%%%%%%%%%%%%%%%%%%%%%%%%%%%%%%%%%%%%%%%%%%%%%%%%%%%%%%%%%%%%%%
\section{Quantum fluctuations}
\label{bog}

Hamiltonian (\ref{ham}) describes a small number of bosonic impurity 
particles immersed in a BEC in a 1D-model with inter-particle
interactions of the contact type.  Assuming that the presence of the
impurity atoms does not perturb the condensate significantly, we may
decompose the bosonic field operator into 
\be
\hat\varphi(x)\approx \sqrt{\rho}+\hat\phi(x),
\label{rozw}
\ee
where $\sqrt{\rho}$ denotes the stationary mean-field solution for a
condensate
wavefunction in the absence of impurities, and where $\hat{\phi}(x)$
describes
the small perturbations of the condensate caused by the impurity (or
impurities).
 We substitute 
(\ref{rozw}) into the Hamiltonian (\ref{ham}) and keep terms that are quadratic in
the $\hat\phi$ operator only. 
Next, we replace the condensate boson-boson interaction term
in the Hamiltonian by
\be
\frac{\lambda_{{\rm BB}}}{2}\rho\left(
\hat\phi\hat\phi+
4\hat\phi^\dagger\hat\phi+
\hat\phi^\dagger\hat\phi^\dagger
\right)
\approx 
3\lambda_{{\rm BB}}\rho\hat\phi^\dagger\hat\phi.
\label{appBEC}
\ee
This approximation implies that we neglect the depletion of the BEC induced
by the interactions between condensate particles.  The BEC-depletion that we compute
is then caused entirely by the interactions with the impurities.  
We expect that this approximation will not greatly affect the impurity physics.
It does, however, modify the description of long wavelength BEC-excitations.  
In the absence of impurities the  approximation (\ref{appBEC}) corresponds 
to the Hartree-Fock approximation.  This approach predicts a gap in the 
BEC-excitation spectrum, contrary to the Goldstone (or Hugenholtz-Pines) theorem,
and different from the the Bogoliubov spectrum \cite{BT}. 
The final effective Hamiltonian in the units (\ref{units}) reads
\bea
\hat H_{\rm eff}&=&\int {\rm d}x\left\{
\hat\phi^\dagger(x)\left[
-\frac{m_{\rm I}\partial^2_x}{2m_{\rm B}}+\alpha \right] \hat\phi(x) 
\right. \cr
&+& \hat\psi^\dagger(x)\left[
-\frac{\partial^2_x}{2}-E \right] \hat\psi(x) \cr
&+& 
\left.
\gamma \left[\hat\phi^\dagger(x)+\hat\phi(x)\right]\hat\psi^\dagger(x)\hat\psi(x)
\right\},
\label{eff}
\eea
where
\bea
\alpha&=& \frac{2\hbar}{\lambda_{\rm IB}^2}\sqrt{\frac{\lambda_{\rm
BB}^3\rho}{m_{\rm B}}},
\cr
\gamma&=& {\rm sign}(\lambda_{\rm IB})\left(\frac{\hbar^2\lambda_{\rm BB}^3m_{\rm I}^2\rho}{\lambda_{\rm
IB}^4m_{\rm B}^3}\right)^{1/8}.
\eea

We treat the resulting Hamiltonian (\ref{eff}) with a  Bogoliubov 
approximation \cite{BT}:
first we solve the mean-field equations, then we construct the quantum
fluctuations
around the mean-field solutions.  To this end, we expand the field operators
as 
\bea
\hat\phi(x) &\approx& \phi_0(x)+\delta\hat\phi(x), \cr
\hat\psi(x) &\approx& \psi_0(x)+\delta\hat\psi(x). 
\label{bog1}
\eea
In zeroth order in the 
$\delta\hat\phi$ and $\delta\hat\psi$ operators one obtains the 
mean field equations,
\bea
\left[-\frac{m_{\rm I}\partial_x^2}{2m_{\rm B}}+\alpha\right]\phi_0(x)
+\gamma|\psi_0(x)|^2&=&0, \cr
\left[-\frac{\partial_x^2}{2}+\gamma\left\{\phi_0(x)+\phi_0^*(x)\right\}
\right]\psi_0(x) &=&E\psi_0(x),
\label{meanf4}
\eea
identical to Eqs.~(\ref{meanf2a},\ref{meanf2b}), written in the
units~(\ref{units}). 

While the mean-field description of coupled atomic-molecular BEC's give
solitonic solutions that are identical to Eq. (\ref{soliton})
\cite{drummondPRL98,drummondPRA04}, the Hamiltonian of
the atomic-molecular BEC's is different from Eq.~(\ref{eff}). The coupled
atomic-molecular BEC Hamiltonian contains terms that convert
molecules into atoms and vice versa, rather than
the interaction term 
$\gamma \left[\hat\phi^\dagger(x)+\hat\phi(x)\right]\hat\psi^\dagger(x)
\hat\psi(x)$. The latter term can be found in models where bosonic 
particles $\hat\psi$  feel a long-range force caused by the exchange of a 
mesonlike particle $\hat\phi$ \cite{drummondPRA04}.

Given that $\psi_{0}(x)$ and $\phi_{0}(x)$ satisfy the mean-field
equations (\ref{meanf4}), the first order terms in the $\delta \hat{\phi}$ and
$\delta \hat{\psi}$ operators cancel.  The second order term gives an
effective Hamiltonian that takes the form
\be
\hat H_{\rm eff}\approx \frac12 \int {\rm d}x
\left(\delta\hat\phi^\dagger,-\delta\hat\phi,\delta\hat\psi^\dagger,
-\delta\hat\psi\right) {\cal L}
\left(\begin{array}{c} 
\delta\hat\phi \\ 
\delta\hat\phi^\dagger \\
\delta\hat\psi \\
\delta\hat\psi^\dagger
\end{array}\right),
\label{effapp}
\ee 
where
\be
{\cal L}=\left(
\begin{array}{cccc}
 h_\phi  &  0 &  \gamma\psi_0^*(x) & \gamma\psi_0(x)  \\
 0  & -h_\phi  & -\gamma\psi_0^*(x) & -\gamma\psi_0(x) \\
 \gamma\psi_0(x)  & \gamma\psi_0(x)  & h_\psi  &  0 \\
 -\gamma\psi_0^*(x)  &  -\gamma\psi_0^*(x) & 0  &  -h_\psi \\
\end{array}
\right),
\label{Lop}
\ee
and
\bea
h_\phi&=& -\frac{m_{\rm I}\partial_x^2}{2m_{\rm B}}+\alpha \cr
h_\psi&=&-\frac{\partial_x^2}{2}+\gamma\left\{\phi_0(x)+\phi_0^*(x)\right\}-E.
\eea
The problem of diagonalizing the Hamiltonian (\ref{effapp}) reduces to 
the problem of diagonalizing the non-Hermitian operator $\cal L$  (i.e.  to solving 
equations of the Bogoliubov-de~Gennes type). 
The $\cal L$--operator possesses two symmetries (similar to the 
symmetries of the original Bogoliubov-de~Gennes equations \cite{castin}),
\bea
u_1{\cal L}u_1&=&-{\cal L}^*, \cr
u_3{\cal L}u_3&=&{\cal L}^\dagger,
\label{symmL}
\eea
where
\bea
u_1=\left(
\begin{array}{c|c}
 \sigma_1 & 0 \\
\hline
 0 & \sigma_1 \\
\end{array}
\right),
&&
u_3=\left(
\begin{array}{c|c}
 \sigma_3 & 0 \\
\hline
 0 & \sigma_3 \\
\end{array}
\right),
\eea
and 
\bea
\sigma_1=\left(\begin{array}{cc}
0&1\\
1&0
\end{array}\right), &&
\sigma_3=\left(\begin{array}{cc}
1&0\\
0&-1
\end{array}\right),
\eea
are the first and third Pauli matrices, respectively. 

Suppose that all eigenvalues of the ${\cal L}$--operator are real. The symmetries  
(\ref{symmL}) imply that if
\be
|\Psi_k^{\rm R}\rangle=
\left(
\begin{array}{c}
|u_k^\phi\rangle  \\
|v_k^\phi\rangle \\
|u_k^\psi\rangle  \\
|v_k^\psi\rangle \\
\end{array}
\right),
\ee
is a right eigenvector of the ${\cal L}$-operator
with eigenvalue $\varepsilon_k$, then 
$|\Psi_k^{\rm L}\rangle=u_3|\Psi_k^{\rm R}\rangle$
is a left
eigenvector of the same eigenvalue $\varepsilon_{k}$, and 
$u_{1} |\Psi^{R\ast}_{k}\rangle$
is a
right eigenvector with eigenvalue $-\varepsilon_{k}$.  Except for the
eigenstates
corresponding to zero eigenvalue, the eigenstates of the ${\cal
L}$--operator
can be divided into two families "+" and "$-$",
\be
\langle\Psi_k^{\rm R}| u_3|\Psi_{k'}^{\rm R}\rangle=\pm \delta_{k,k'}.
\ee
We apply the Bogoliubov transformation,
\be
\left(\begin{array}{c} 
\delta\hat\phi(x) \\ 
\delta\hat\phi^\dagger(x) \\
\delta\hat\psi(x) \\
\delta\hat\psi^\dagger(x)
\end{array}\right)=
\sum_{k\in"+"}\left[
\hat b_k \Psi_k^{\rm R}(x)+\hat b_k^\dagger u_1\Psi_k^{\rm R*}(x)
\right],
\ee
with quasi-particle operators that can be written as
\bea
\hat b_k^\dagger&=& - \langle v_k^{\phi*} | \hat\phi \rangle + \langle
u_k^{\phi*} | \hat\phi^\dagger \rangle
- \langle v_k^{\psi*} | \hat\psi \rangle + \langle u_k^{\psi*} | \hat\psi^\dagger \rangle,
\cr
\hat b_k&=& \langle u_k^\phi | \hat\phi \rangle - \langle v_k^\phi | \hat\phi^\dagger \rangle
+ \langle u_k^\psi | \hat\psi \rangle - \langle v_k^\psi | \hat\psi^\dagger \rangle, 
\label{boper}
\eea
and that fulfill the bosonic commutation relation
$[ \hat{b}_{k}, \hat{b}^{\dagger}_{k'} ] = \delta_{k,k'}$.
This transformation gives an effective Hamiltonian that is
diagonal
\be
\hat H_{\rm eff}\approx\sum_{k\in"+"} \varepsilon_k \hat b_{k}^\dagger\hat b_k.
\label{efffin}
\ee

To obtain the energy eigenvalues and eigenstates of the effective
Hamiltonian (\ref{effapp}), one has to solve the mean-field equations
(\ref{meanf4}), then diagonalize the operator (\ref{Lop}).  
For a specific value of
the average BEC-density, given by Eq.~(\ref{rhos}), we obtain the solitonic
solution (\ref{soliton}) of the mean-field equations.  In that case, the
eigenstates of the ${\cal L}$--operator corresponding to zero-eigenvalue
take on the form
\bea
\left(
\begin{array}{c}
0  \\
0 \\
\cosh^{-2}\left(\sqrt{\frac{M}{6}}x\right)  \\
-\cosh^{-2}\left(\sqrt{\frac{M}{6}}x\right) \\
\end{array}
\right), \cr
\left(
\begin{array}{c}
-\sqrt{\frac{m_{\rm B}}{2m_{\rm I}}}  \\
-\sqrt{\frac{m_{\rm B}}{2m_{\rm I}}} \\
{\rm sign}(\lambda_{\rm IB})  \\
{\rm sign}(\lambda_{\rm IB}) \\
\end{array}
\right)\partial_x\cosh^{-2}\left(\sqrt{\frac{M}{6}}x\right),
\eea
where the first eigenvector corresponds to the breaking of the U(1)
symmetry in the BEC-Bogoliubov theory \cite{castin}, while the second
eigenvector corresponds to the breaking of the translational
symmetry, indicating that the translation of the soliton costs no energy.
There is another zero-momentum eigenstate which has non-zero
eigenvalue.  This eigenvalue and its eigenvector take on simple analytical
forms,
\bea
\varepsilon&=&\frac{Mm_{\rm I}}{3m_{\rm B}}, \cr
\Psi^{\rm R}(x)&=&\left(
\begin{array}{c}
{\rm sign}(\lambda_{\rm IB})  \\
0 \\
\frac32\sqrt{\frac{m_{\rm B}}{2m_{\rm I}}}\cosh^{-2}\left(\sqrt{\frac{M}{6}}x\right)  \\
-\frac32\sqrt{\frac{m_{\rm B}}{2m_{\rm I}}}\cosh^{-2}\left(\sqrt{\frac{M}{6}}x\right) \\
\end{array}
\right)\frac{1}{\sqrt{L}},
\label{nontrm}
\eea
where $L$ is the size of the 1D-box.  Other eigenstates can be found
numerically.  In the next section, we present numerical results for
realistic parameter values.

%%%%%%%%%%%%%%%%%%%%%%%%%%%%%%%%%%%%%%%%%%%%%%%%%%%%%%%%%%%%%%%%%%%%%
\section{Numerical results}
\label{num}

\begin{figure}
\centering
\includegraphics*[width=8.6cm]{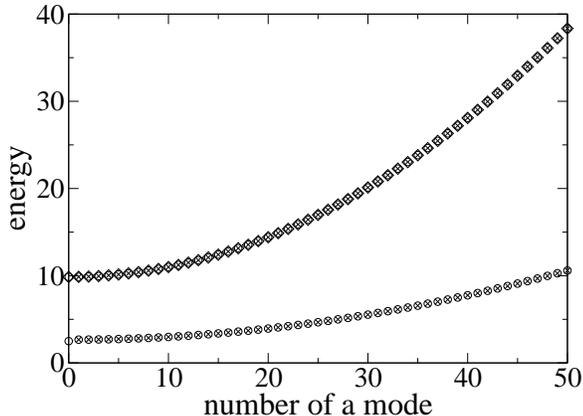}
\caption{  
Excitation spectrum of the quasi 1D $^{23}$Na BEC with eight
$^{85}$Rb atom impurities, in the energy units of Eq.~(\ref{units}).
The upper branch corresponds to
BEC-particle like excitations, while the impurities remain localized.
This branch coincides with the Hartree-Fock spectrum of 
the BEC without the impurity atoms (diamonds) given by 
Eq.~({\protect \ref{hf}}).
The lower branch corresponds to excitations that delocalize the
impurity atoms (except for the lowest energy level, as can be
seen in Fig.\ref{modes1}).  Circles denote even solutions and crosses show
solutions of odd symmetry.
}
\label{spec}
\end{figure}

We consider eight impurity $^{85}$Rb atoms, i.e. $M=8$, immersed in a BEC  
of $^{23}$Na atoms in an elongated trap that is a box in the axial direction 
while in the transverse directions there is a harmonic trap of frequency $\omega_\perp$. The
transverse $\omega_{\perp}$--confinement is so strong that only
the ground states of the transverse degrees of freedom are relevant.
In this quasi 1D-regime, the system is effectively one dimensional
and confined by a 1 D-box potential.  The coupling constants of
the Hamiltonian (\ref{ham}) read
\bea
\lambda_{\rm IB}&=&2\pi \hbar^2 a_{\rm IB}\left(\frac{1}{m_{\rm I}}+
\frac{1}{m_{\rm B}}\right)
\frac{1}{\pi(\sigma_{\rm B}^2+\sigma_{\rm I}^2)}  \cr
&=&2\hbar\omega_\perp a_{\rm IB}, \cr
\lambda_{\rm BB}&=&\frac{4\pi \hbar^2 a_{\rm BB}}{m_{\rm B}}\frac{1}{2\pi\sigma_{\rm
B}^2}=2\hbar\omega_\perp a_{\rm BB}, 
\eea
where the $s$-wave scattering lengths $a_{\rm BB}=3.4$~nm and 
$a_{\rm IB}=16.7$~nm \cite{bigelow}. The 
$\sigma_{\rm B}=\sqrt{\hbar/m_{\rm B}\omega_\perp}$ and 
$\sigma_{\rm I}=\sqrt{\hbar/m_{\rm I}\omega_\perp}$ lengths represent the
ground state extents of a single BEC-boson particle and of a single
impurity atom confined by the two-dimensional harmonic trap in the
transverse direction with frequency $\omega_{\perp}$.  We take the
transverse trapping frequency to be equal to $\omega_{\perp}
= 2 \pi \times 500$~Hz.
Assuming that the size of the box in the longitudinal direction
is equal to $100 \mu$m (in units (\ref{units}), this length corresponds to
a box size $L \approx 80$), we confine $11000$ atoms in the
BEC.  These parameters correspond to the specific value of the 
condensate density (\ref{rhos}) and are well within the experimental range 
\cite{raizen}.

\begin{figure}
\centering
\includegraphics*[width=8.6cm]{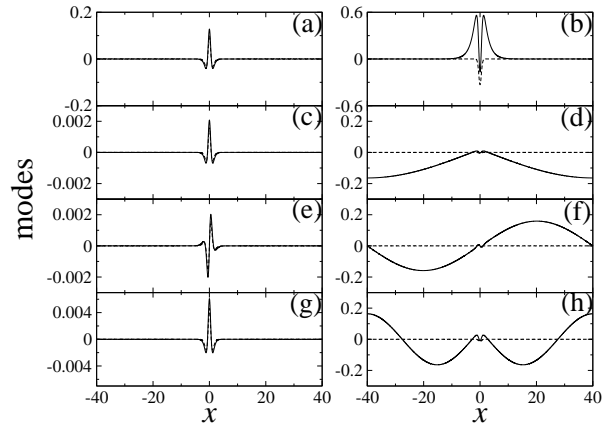}
\caption{  
The eigenstates of the $\cal L$-operator, Eq.~({\protect \ref{Lop}}), 
corresponding to the lower energy branch of the Hamiltonian ({\protect
\ref{efffin}}), see Fig.~{\protect \ref{spec}}. Left column is related to 
condensate modes, i.e. $u_k^{\phi}$ (solid lines) and $v_k^{\phi}$ (dashed 
lines --- hardly visible behind the solid lines), while the right column is related to 
impurities modes, i.e. $u_k^{\psi}$ (solid lines) 
and $v_k^{\psi}$ (dashed lines). There are four lowest energy eigenstates
presented in the figure, i.e. (a)-(b) is the lowest one, (c)-(d), (e)-(f) and 
(g)-(h) are shown in order of increasing energy.  The
states describe the excitation of eight self-localized $^{85}$Rb
atoms embedded in a quasi 1D $^{23}$Na BEC.
}
\label{modes1}
\end{figure}

\begin{figure}
\centering
\includegraphics*[width=8.6cm]{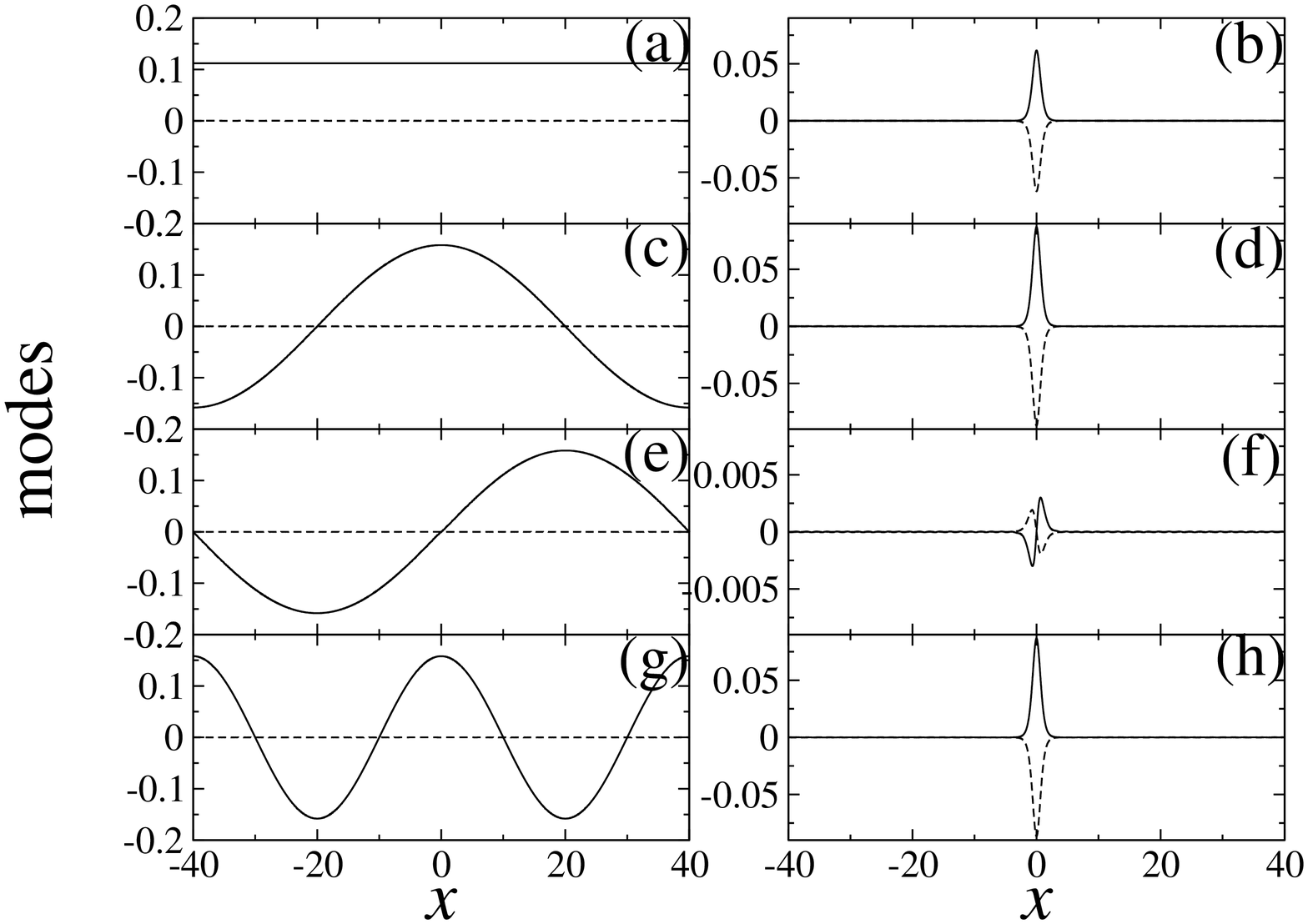}
\caption{  
The same as in Fig.~{\protect \ref{modes1}} but for
the upper energy branch of the Hamiltonian ({\protect
\ref{efffin}}), see Fig.~{\protect \ref{spec}}. The modes shown
in panels (a)-(b) correspond to the solution given by (\ref{nontrm}).
}
\label{modes2}
\end{figure}

Figure~\ref{spec}  shows the excitation spectra that were calculated 
numerically with
periodic boundary conditions.  Note that the dispersion shows two
branches.  Each excitation of the lower branch delocalizes an impurity
atom, except for the very lowest level, which corresponds to an
excitation that leaves the impurity atom localized, as shown in Fig.~\ref{modes1}.
The upper branch corresponds to particle-like BEC excitations in
which the impurity remains localized, as shown in Fig.~\ref{modes2}.  The
highest curves of Fig.~\ref{modes2} show the mode of Eq.~(\ref{nontrm}).  
In Fig.~\ref{spec} we have also plotted the BEC-spectrum
calculated within the Hartree-Fock approximation \cite{BT},   
\be
\varepsilon^{\rm HF}_n=\frac{m_{\rm I}}{m_{\rm B}}\left[
\frac{2\pi^2}{L^2}n^2+\frac{M}{3}
\right],
\label{hf}
\ee
where $n=0,1,\ldots$. Actually, to obtain (\ref{hf}) we have used the 
Hamiltonian (\ref{eff}) in which we neglected terms involving the $\hat\psi(x)$ 
operator. The Hartree-Fock spectrum (\ref{hf}) provides the relevant
comparison for the model since the terms neglected in Eq.(\ref{appBEC})
correspond to the Hartree-Fock description of the single BEC.
Figure~\ref{spec} shows that the upper energy branch is nearly identical
to the Hartree-Fock BEC-spectrum in the absence of impurities. 
The lower energy branch describes impurity excitations with negligible
excitation of the BEC, see  Fig.~\ref{modes1}).  
We interpret the energy gap in the lower energy branch (as the wavenumber tends
to zero) as the minimal energy needed to break the self-localization bond of
the many-body BEC-impurity system.  This gap is then a characteristic feature
of impurity self-localization.  Its detection by means of Bragg spectroscopy
\cite{bragg}, for instance, can serve as a smoking gun signal for the observation of
BEC-impurity self-localization. 

\begin{figure}
\centering
\includegraphics*[width=8.6cm]{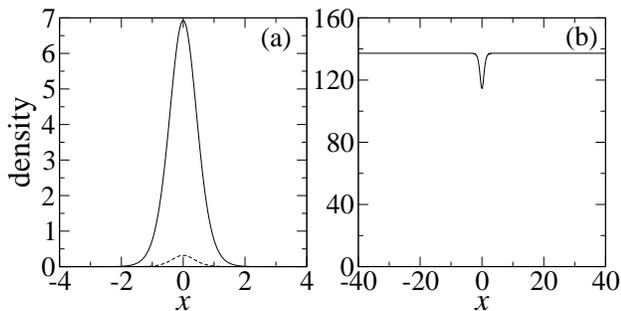}
\caption{  
Panel (a): solid line denotes the density of the impurity atoms,
$\psi_0^2(x)$, calculated within the product state ansatz, dashed line is 
the correction to this density, i.e.
$\langle 0|\delta\hat\psi^\dagger(x)\delta\hat\psi(x)|0\rangle=
\sum_{k}\left[v^{\psi}_k(x)\right]^2$, obtained within the formalism that goes
beyond the product state approximation. Panel (b): BEC density given by 
Eq.~({\protect \ref{BECden}}).
The results are related to eight $^{85}$Rb atoms immersed in a BEC of $^{23}$Na 
atoms. The length is given in the units (\ref{units}). 
}
\label{depl}
\end{figure}

\begin{figure}
\centering
\includegraphics*[width=8.6cm]{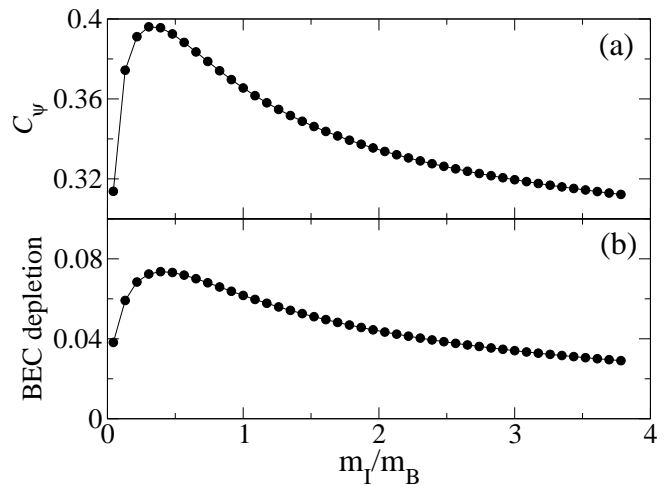}
\caption{  
The integral (\ref{integ}) [panel(a)] and condensate depletion [panel (b)] 
as a function of the impurity mass $m_{\rm I}$. Note that the maximal values 
of the $C_\psi$ and condensate depletion correspond to 
$m_{\rm I}/m_{\rm B}\approx 0.5$. 
}
\label{ratio}
\end{figure}

The ground state of the system is the Bogoliubov vacuum state $|0\rangle$, 
i.e. the state annihilated by all $\hat b_k$ operators (\ref{boper}). 
We can estimate the quantum fluctuation corrections to the product state
by calculating the density of the impurity atoms,
\bea
\langle 0|\hat\psi^\dagger(x)\hat\psi(x)|0\rangle&=&\psi_0^2(x)+
\langle 0|\delta\hat\psi^\dagger(x)\delta\hat\psi(x)|0\rangle \cr
&=&\psi_0^2(x)+\sum_{k\in "+"}\left[ v^{\psi}_k(x)\right]^2.
\label{deplpsi}
\eea
In Fig.~\ref{depl} we plot $\psi_0^2(x)$ and 
$\sum_{k}\left[v^{\psi}_k(x)\right]^2$.
The quantum fluctuation contribution to the impurity density is
\be
C_\psi=\int {\rm d}x \sum_{k\in "+"}\left[v^{\psi}_k(x)\right]^2=0.31,
\label{integ}
\ee
as compared to the value $8$ for the integral of $\psi_{0}^{2}(x)$, which
shows that the fluctuation corrections to the product state are not insignificant. 
In the case of a single impurity, 
i.e. $M=1$, keeping the other parameters fixed, $C_\psi=0.30$ 
and the fluctuations corrections turn out to be more dramatic.
The condensate density (see Fig.~\ref{depl}) reads,
\bea
\langle 0|\hat\varphi^\dagger(x)\hat\varphi(x)|0\rangle&=&
\left[\sqrt{x_0\rho}+\phi_0(x)\right]^2 \cr
&+&\langle 0|\delta\hat\phi^\dagger(x)\delta\hat\phi(x)|0\rangle \cr
&\approx&
x_0\rho+2\sqrt{x_0\rho}\phi_0(x),
\label{BECden}
\eea
where we have introduced approximations due to the facts that
$\phi_0(x)/\sqrt{x_0\rho}\le 0.083$ [that justifies also the
linearization (\ref{lin1}) or (\ref{rozw})] and 
$\langle 0|\delta\hat\phi^\dagger(0)\delta\hat\phi(0)|0\rangle/\phi_0^2(0)
\approx 0.03$.

Decreasing the mass of the impurity atom, while keeping all other 
parameters fixed, the integral (\ref{integ}) and the BEC depletion increase
and reach their maximal values for $m_{\rm I}/m_{\rm B}\approx 0.5$, see Fig.~\ref{ratio}.
This shows that when the impurity mass is half that of the BEC-bosons, the product
state approximation needs the largest corrections.
Qualitatively, this behavior agrees with the time scale separation argument,
which predicts that the large and small mass ratio regimes are well described 
by a product state  \cite{landau}.

%%%%%%%%%%%%%%%%%%%%%%%%%%%%%%%%%%%%%%%%%%%%%%%%%%%%%%%%%%%%%%%%%%%%%
\section{Conclusions}
\label{concl}

We have considered the self-localization of neutral, bosonic impurity atoms embedded
in a dilute gas Bose-Einstein condensate in a 1D-model.  We have
analyzed the system within a Bogoliubov formalism that describes the
quantum fluctuations around the strong coupling product state approximation
previously made in the cold atom literature \cite{eddy,blume}.  
Our description gives an excitation spectrum that consists of two branches.  
The lower energy branch corresponds to excitations that delocalize the impurity 
atoms (with the exception of the lowest energy excitation).  
The higher energy branch corresponds to BEC-particle like excitations
(and coincides, in our approximation, with the Hartree-Fock spectrum of the BEC in the absence of 
the impurity atoms) while the impurities remain self-localized. 
The energy gap in the lower energy branch (i.e. the branch 
that corresponds to the excitation of impurity atoms with negligible 
excitation of the BEC) suggests a spectroscopic means
(Bragg spectroscopy \cite{bragg}) for detecting BEC-impurity self-localization.

The parameters of our calculation are well within the experimental range.
An extension of this approach to the three-dimensional BEC-impurity
system appear straightforward.
One can extend the formalism to account for the presence of a larger
number of impurity atoms by
including a nonlinear term in the Hamiltonian (\ref{ham}) 
to account for interactions among the impurity 
atoms. Also the depletion of the BEC  
caused by the interactions among condensate particles (neglected in the
present calculations) can be included.

Furthermore, our calculations reveal the close analogy of 1D BEC-impurity
self-localization with parametric solitons known in 
nonlinear optics \cite{drummond96,karpierz}, 
in the Schr\"odinger-Newton model \cite{ruffini,diosi,penrose} and in the coupled 
atomic-molecular BEC system \cite{drummondPRL98,drummondPRA04}. The problem of 
impurity atoms immersed in an atomic BEC offers intriguing opportunities for 
experimentally realizing the phenomena predicted in these fields.

%%%%%%%%%%%%%%%%%%%%%%%%%%%%%%%%%%%%%%%%%%%%%%%%%%%%%%%%%%%%%%%%%%%%%
\section*{ Acknowledgments }  
We are grateful to Jacek Dziarmaga and Zbyszek Karkuszewski for 
the helpful discussion.
One of the authors (K.S.) acknowledges support from the Fulbright
Scholar Program and by the KBN grant PBZ-MIN-008/P03/2030.  This
work was funded, in part, by the LDRD Los Alamos program.

%%%%%%%%%%%%%%%%%%%%%%%%%%%%%%%%%%%%%%%%%%%%%%%%%%%%%%%%%%%%%%%%%%%%%

\end{document}